\documentclass[aps,prl,twocolumn,groupedaddress,floatfix,showpacs]{revtex4}

\usepackage{bm} \usepackage{graphicx}
\usepackage{epsfig}
\usepackage{subeqn}

\begin{document}

\title{Radial and angular rotons in trapped dipolar gases}
\author{Shai Ronen} 
\affiliation{JILA and Department of Physics, University of Colorado, Boulder, CO 80309-0440}
\author{Daniele C. E. Bortolotti} 
\affiliation{JILA and Department of Physics, University of Colorado, Boulder, CO
80309-0440} 
\affiliation{LENS and Dipartimento di Fisica, Universit\'{a} di Firenze, Sesto Fiorentino, Italy}
\author{John L. Bohn} 
\affiliation{JILA, NIST, and Department of Physics, University of Colorado,
 Boulder, CO 80309-0440 }
\email{bohn@murphy.colorado.edu}
\date{\today}
\begin{abstract}
We study Bose-Einstein condensates with purely dipolar interactions in
oblate (pancake) traps. We find that the condensate always becomes
unstable to collapse when the number of particles is sufficiently
large. We analyze the instability, and find that it is the trapped-gas
analogue of the ``roton-maxon'' instability previously reported for a
gas that is unconfined in two dimensions.  In addition, we find that
under certain circumstances, the condensate wave function attains a
biconcave shape, with its maximum density away from the center of the
gas. These biconcave condensates become unstable due to azimuthl
excitation - an angular roton.
\end{abstract}


\maketitle


The realization of a Bose-Einstein condensate (BEC) of $^{52}$Cr
\cite{Griesmaier05,Stuhler05,Griesmaier06} marks a major development
in degenerate quantum gases. The inter-particle interaction via
magnetic dipoles in this BEC is much larger than that in alkali atoms,
and leads to an observable change in the shape of the condensate. In
these experiments, the dipolar interaction energy was about 15\% of
the short range interaction energy given by the scattering length. It
may be expected that advances in utilizing Feshbach resonances
\cite{Werner05} to tune the scattering length to zero, will lead in
the near future to condensates dominated by pure dipolar interactions.
A rich array of intriguing phenomena have already been predicted to
occur in such a gas \cite{Goral00,Yi00,Santos00,Yi01,Yi02,Goral02,
Baranov02,Santos03,ODell04,Eberlein05,Nho05,Uwe06,Bortolotti06,Ronen06,Ronen06a}.

It is widely agreed that the structure and stability of a dipolar gas
will depend critically on the interplay between three factors: the
anisotropic dipole-dipole interaction between particles; the
anisotropic external potential used to confine the particles; and the
number of particles. Still, the details of this interplay remain thus
far fragmentary. We take the standard scenario, where the dipoles are
assumed to be polarized along a field whose direction defines the
laboratory $z$ axis. They are confined by a harmonic trap with
cylindrical symmetry about this axis, defined by radial and axial
angular frequencies $\omega_{\rho}$ and $\omega_z$. It has been
realized that, because the dipoles would prefer to align in a
head-to-tail orientation, the gas elongates in the $z$ direction and
need not possess the same aspect ratio $\lambda \equiv
\omega_z/\omega_{\rho}$ as the trap \cite{Yi00,Yi01}. If this is so,
then ultimately the attractive interaction between dipoles should
become so large as to initiate a macroscopic collapse as the number of
dipoles is increased. In fact, in the large-number limit, the gas
approaches the Thomas-Fermi regime, in which a dipolar gas is
inherently unstable \cite{Eberlein05}.

The circumstances under which this collapse occurs has been the
subject of much discussion. It was argued in Ref. \cite{Santos00}
that for oblate traps with $\lambda > 5.4$, the gas is completely
stable, since the trap can overcome the tendency of the gas to
elongate along $z$. However, Ref. \cite{Yi01} found unstable
condensates even for $\lambda = 7$. Taking it a step further,
Ref. \cite{Santos03} found instabilities even in the $\lambda
\rightarrow \infty$ limit, by treating a gas that is confined in the
$z$ direction, but completely free to move in the transverse plane.
The instability was refereed to as a ``roton'' instability, although
the microscopic physics is quite different from the roton-maxon
physics in superfluid He. Ref.~\cite{Uwe06} examined the
quasi-two-dimensional limit of this problem. Roton-like dispersion
curves and supersolid-like phases had earlier been identified in
quasi-1D gases, where the dipole dipole interaction is induced by
strong laser light \cite{Giovanazzi02a,ODell03,Mazets04,Kurizki04}.

In any event, the details of the stability of a pure dipolar gas
remain an open question. In this Letter, we re-visit this issue,
applying a new efficient algorithm \cite{Ronen06a}. We determine that
at any finite value of $\lambda$, the condensate will decay if
sufficiently many dipoles are added. The route toward instability
will be shown to exhibit physics similar to the roton mechanism, but
adapted to a finite trap environment. Along the way, we demonstrate
that stable condensates can have an unusual density profile, where the
maximum of density is {\it not} in the center of the gas.

The dynamics of the condensate wave function
$\psi({\bm{r},t})$ is described by the time-dependent Gross-Pitaevskii
(GP) equation:
\begin{widetext}
\begin{eqnarray}
\lefteqn{i \hbar \frac{\partial\psi(\bm{r},t)}{\partial t}
=\Big[-\frac{\hbar^{2}}{2M}\nabla^{2}+\frac{\omega_{\rho}^{2}}{2}
(\rho^{2}+\lambda^{2}z^{2})+} \nonumber \\ 
& & (N-1)\frac{4\pi\hbar^2 a}{M}|\psi(\bm{r},t)|^2
+(N-1) \int d\bm{r'}V_d(\bm{r}-\bm{r'})
|\psi(\bm{r'},t)|^2\Big]\psi(\bm{r},t),
\label{gpe}
\end{eqnarray}
\end{widetext}
where $M$ is the particle mass and the wave function $\psi$ is
normalized to unit norm. The coupling constant for the short-range
interaction is proportional to the scattering length $a$, although we
set $a=0$ in the following. The dipole-dipole interaction is given by
$V_d(\bm{r})=d^2(1-3\cos^{2}\theta)/r^3$, with $\theta$ being the
angle between the vector $\bm{r}$ and the $z$ axis. We also define a
dimensionless dipolar interaction parameter, $D=(N-1)M
d^2/(\hbar^{2}a_{ho})$, where $a_{ho}=\sqrt{ \frac{\hbar} {M
\omega_{\rho} } } $ denotes the transverse harmonic oscillator
length. It is convenient to think of increasing $D$ as equivalent to
increasing the number of dipoles, for a fixed trap geometry.

\begin{figure}
\vspace{-0.2in}
\resizebox{3.1in}{!}{\includegraphics{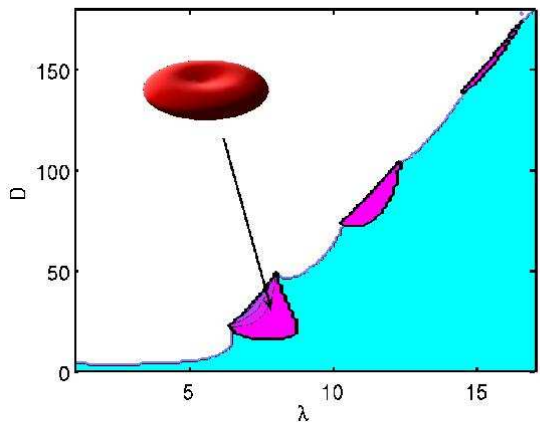}}
\vspace{-0.1in}
\caption{(Color online) Stability diagram of a dipolar condensate in a
trap, as a function of the trap aspect ratio
$\lambda=\omega_z/\omega_{\rho}$ and the dipolar interaction parameter
$D$. Shadowed areas are stable, while white unstable against
collapse. In the darker, isolated areas, we find biconcave condensates
(illustrated with iso-density surface plot at the top right corner)
whose maximal density is not at the center. The contours in the
biconcave regions indicate the ratio of the central density to the
maximal density, with darker areas having a smaller ratio. The contour
intervals are 10\%, and the minimum ratio obtained is 70\%.
\label{fig:phase1}}
\end{figure}

\begin{figure*}
\vspace{-0.2in}
\resizebox{6.7in}{3.3in}{\includegraphics{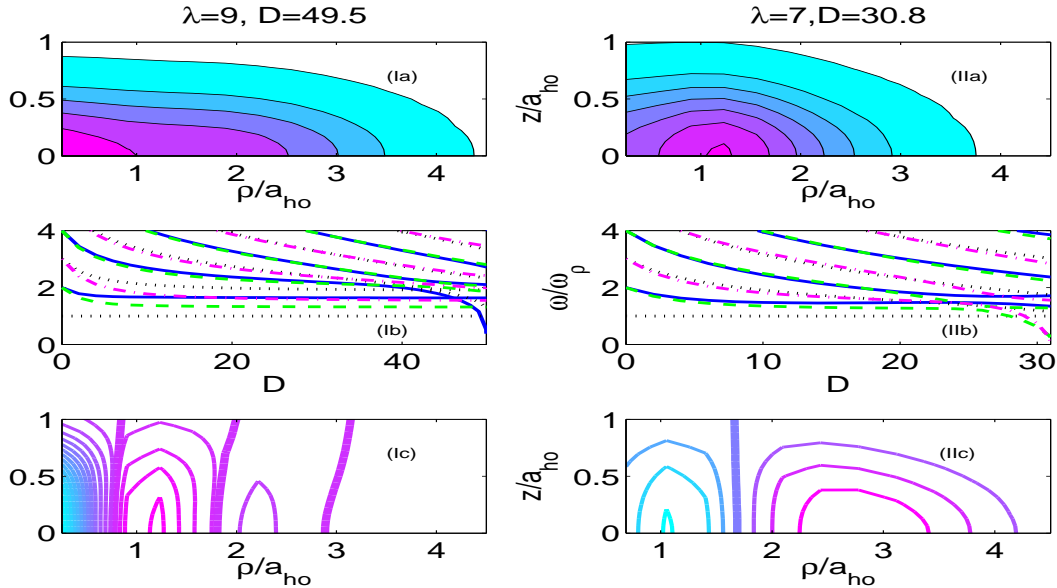}}
\vspace{-0.3in}
\caption{(Color online) (Ia) Contour plot of the ground state density
for a dipolar condensate in a trap with aspect ratio $\lambda=9$ and
dipolar interaction strength $D=49.5$ (close to instability).. (IIa)
Condensate density contour plot for $\lambda=7$ and $D=30.8$. Note
the unusual shape, with the maximum of the density occurring away from
the center of the trap, along a ring (the plot is a vertical
cross-section through the cylindrically symmetric condensate). The
3-dimensional shape is biconcave (inset of Fig.~\ref{fig:phase1}).
(Ib) Excitation spectrum as a function of $D$ for $\lambda=9$, and (IIb)
for $\lambda$=7. Solid lines are excitation with angular quantum number
$m=0$, dotted lines $m=1$, dashed lines $m=2$ and dot-dashed lines
$m=3$. (Ic) The density perturbation (see text)of the lowest
excitation mode for the ground state of (Ia), and the corresponding
excitation (IIc) for the ground states of (IIa). The heavy lines are
nodal lines (actually surfaces in 3D) of the excitation.
\label{fig:panel}}
\end{figure*}

A stable condensate exists when there is a stationary solution of
Eq.~(\ref{gpe}) which is stable to small perturbations. Based on this
criterion, we have constructed the stability diagram in
Fig.~\ref{fig:phase1}, as a function of $D$ and the trap aspect ratio
$\lambda = \omega_z / \omega_{\rho}$. In the figure, the shaded and
white areas denote parameter ranges for which the condensate is stable
or unstable, respectively. In general, the more pancake-like the trap
becomes (larger $\lambda$), the more dipoles are required to make the
condensate unstable.

Remarkably, there appear regions in parameter space where the
condensate obtains its maximum density away from the center of the
trap. These are the darker shaded areas in Fig.~\ref{fig:phase1}. The
local minimum of the density in the center gives the condensate a
biconcave shape, resembling that of a red blood cell (a surface of
constant density is illustrated at the top left corner of
Fig.~\ref{fig:phase1}). A density contour plot of such biconcave
condensate is shown in Fig.~\ref{fig:panel}(IIa), with the parameters
$\lambda=7$ and $D=30.8$. These structures appear in isolated regions
of the parameter space, and in particular, only for certain aspect
ratios in the vicinity of $\lambda\approx 7,11,15,19...$.
There seems to be a repeated pattern which probably continues to
larger values of $\lambda$ (although this was not
calculated). However, the parameter space area of these regions
becomes increasingly small with larger $\lambda$, and is already
negligible for the fourth region (not shown in
Fig.~\ref{fig:phase1}). In between the biconcave regions, we find
``normal'' condensates with maximum density in the center,
Fig.~\ref{fig:panel}(Ia). Their density profiles in fact have a fairly
sharp peak in the center, as compared to a Gaussian shape.  

To verify the existence of biconcave structures, we solve
Eq.~(\ref{gpe}) numerically both with a 3D algorithm \cite{Goral02}
and with a 2D algorithm that exploits the cylindrical symmetry
\cite{Ronen06a}, carefully converging both the grid size and
resolution. We have also carried out a variational calculation, in
which the condensate wave function is taken to be a linear combination
of two harmonic oscillator wave functions. The first is a simple
Gaussian, and the second is the same Gaussian multiplied by
$(H_2(x)+H_2(y))$, where $H_2$ is the Hermite polynomial of order 2,
and $(x,y)$ are the coordinates perpendicular to the trap
axis. Minimizing the variational energy analytically, we find
biconcave solutions with a large component of the second wave
function, similar in shape to the numerical ones. The exact parameters
for appearance of variational biconcave condensates are somewhat
different from the numerical ones, and the variational calculation
gives a continuous parameter region with biconcave condensates, rather
than isolated regions as we find numerically. This is probably due to
the oversimplified nature of the variational ansatz.

It is perhaps not too surprising that a dipolar condensate could
form with the density pushed to the outer rim. After all, the dipoles
exert long-range, repulsive forces on one another, and flee to the
surface, much as free charges in a conducting material do. What is 
somewhat more mysterious is why the dipoles do not {\it always} behave 
this way, but rather only for certain well-delineated islands in the 
stability diagram.

Condensates with normal and biconcave shapes decay quite differently
as the instability boundary is crossed. To examine the mechanism
causing the collapse of condensates, we have studied collective
excitations in the Bogoliubov-De Gennes (BdG) approximation. Let us
first consider the normal case, where the condensate density is a
maximum in the center of the trap. An example of this density profile
is shown in Fig.\ref{fig:panel}Ia. The excitation spectrum for this
case is shown in Fig.\ref{fig:panel}(Ib). The instability of the
condensate at $D=50.0$ coincides with an excited state mode ``going
soft,'' i.e., tending to zero excitation frequency. In this case the
excited state has zero projection of angular momentum around the $z$
axis, $m=0$. The excitation consists of a radial nodal pattern, as we
show in Fig.~\ref{fig:panel}(Ic) by plotting contours of the eigenmode
density $\delta n=u+v$, where $u,v$ are the usual BdG functions
\cite{Ronen06a}. Thus when this condensate becomes unstable, the
instability is due to a modulation of the condensate density in the
radial direction.

A similar modulation was already considered in the roton
instability studied for the infinite-pancake trap \cite{Santos03}. In
that case, there is only a trap potential in the $z$ direction, and
motion is free in the $(x,y)$ plane, so that the excitations have a
definite transverse momentum $q$ in this plane. The instability is
expected to occur when this momentum approaches the value $q \approx
\hbar/l_z$, where $l_z=\sqrt{\hbar / M \omega_{z}}$ is the axial
confinement length. In this case, excitations in the quasi-2D trap
begin to acquire a 3D character and the interparticle repulsion is
reduced. For sufficiently many dipoles, the attraction can again
overwhelm the condensate and initiate a collapse. Here the
instability is a density-wave modulation $\sin(k x)$ with
characteristic wavelength $2 \pi /k \sim 2 \pi l_z$.

For a gas confined in all three dimensions, the same physics appears.
For aspect ratio $\lambda=9$, the axial confinement length is $l_z =
a_{ho}/3$ (recall that the radial oscillator length defines our
unit of length). Thus the excited state should posses density-wave
oscillations with a wavelength $2 \pi l_z \sim 2 a_{\rm ho}$, or nodes
at intervals $\sim 1 a_{\rm ho}$. The excitation plotted in
Fig.~\ref{fig:panel}(Ic) indeed has radial nodes on approximately
this scale. In this sense, the confined gas exhibits what might be
termed a ``radial roton'' instability.

By contrast, for trap aspect ratios such as $\lambda=7$, where
biconcave condensates appear [Fig.~\ref{fig:panel}(IIa)], the lowest
excitation mode near the instability boundary has angular momentum
projection $m>0$. For example, $m=3$ for the case shown
[Fig.~\ref{fig:panel}(IIb)], meaning that the excited mode exhibits an
azimuthal dependence proportional to $\sin (3\phi)$. Therefore, with
increasing number of particles, these condensates collapse due to
density modulations in the angular coordinate, which spontaneously
break the cylindrical symmetry. Biconcave condensates thus decay to a
kind of ``angular roton'' instability in the trap. [There is also one
radial nodal surface, as seen in the $(\rho,z)$ cross-section in
Fig.~\ref{fig:panel}(IIc).]. The appearance of an angular roton when
the ground state has biconcave shape may be understood in light of the
fact that the maximum density of such a condensate lies a long a
ring. The instability may thus also be described as buckling along
this ring.

In general, the number of the nodal surfaces of the soft mode near the
instability increases with $\lambda$: for radial rotons, from 3 for
$\lambda=9$ to 7 for $\lambda=17$. For angular rotons, we find, for
$\lambda=6.6$, $m=2$ and one radial node; for $\lambda=7$, $m=3$ and
one radial node; and for $\lambda=12$, $m=2$ and 4 radial nodes.

\begin{figure}
\resizebox{3.0in}{!}{\includegraphics{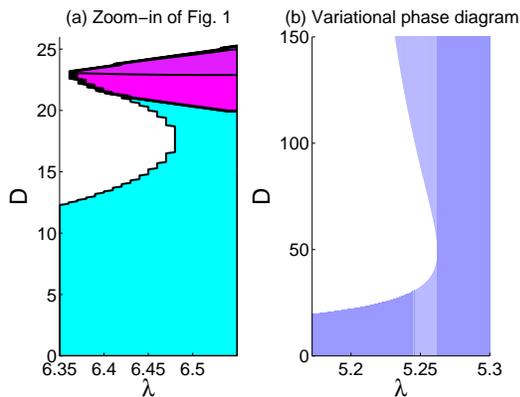}}
\vspace{-0.2in}
\caption{(Color online) (a) Detail of Fig.~\ref{fig:phase1} showing double
stability regions. The coarseness of the contour line is due to
the resolution of our sampling of the parameters space. (b) Part of a
stability diagram using a Gaussian variational method (there is no
area corresponding to biconcave condensates, since these cannot be
described by a Gaussian).
\label{fig:phasezoom}}
\end{figure}

Finally, we point out an unusual feature in the stability diagram,
shown in Fig.~\ref{fig:phasezoom}. Namely, for a small range of aspect
ratios $6.36<\lambda<6.5$, the boundary between stable and unstable
condensates bends back on itself, so that there are two distinct
stability regions as $D$ is varied. This feature is also reproduced
qualitatively by a variational estimate using a single Gaussian
[\ref{fig:phasezoom}(b)]. According to the variational solution, we
find that as $D$ increases, the kinetic energy plays a role in {\it
de-stabilizing} the condensate. Kinetic energy leads to a tendency to
expand, but this tendency will be greater in the more tightly confined
$z$ direction. In this case, more dipoles can pile up along the axis,
where the attraction between dipoles leads to instability. As $D$
increases further, the dipole-dipole interaction becomes more
important, and the effects of kinetic energy are mitigated. Finally,
in the large-$D$ limit, the Thomas-Fermi regime is reached, and
complete instability occurs \cite{Eberlein05}.  Evidently this
dual-stability region depends an a delicate balance between the aspect
ratio of the trap and the dipole-dipole interaction, since it occurs
only over a small region of parameter space.

In conclusion, we have described the stability diagram for dipolar
condensates in pancake traps, found new structured, biconcave
condensates, and an excitation spectrum which exhibits a discrete
roton-maxon. The soft mode of biconcave condensates near collapse
has angular momentum quantum number $m>0$, i.e giving rise to
azimuthal oscillations. This exotic behavior is another
demonstration of the richness of the physics of dipolar condensates.

We are grateful to D. O'Dell for insightful discussions. SR
gratefully acknowledges financial support from the United
States-Israel Educational Foundation (Fulbright Program); DCEB and JLB
from the DOE and the Keck Foundation. 
\bibliography{biblo}
\end{document}